\documentclass[aps, prl, superscriptaddress, twocolumn, floatfix, 10pt]{revtex4-1}
\usepackage{amsmath,amssymb,gensymb,graphicx,color}
\usepackage{hyperref}

\newcommand{\ii}{\mathrm{i}}

\newcommand{\ket}[1]{|#1 \protect\rangle} 

\newcommand{\appref}[1]{Appendix~\ref{#1}}

\newcommand{\refcite}[1]{Ref.\,\cite{#1}}
\newcommand{\eqnref}[1]{Eq.\,(\ref{#1})}
\newcommand{\figref}[1]{Fig.\,\ref{#1}}
\newcommand{\sfigref}[2]{Fig.\,\hyperref[#1]{\ref{#1}(#2)}}

\newcommand{\unit}[1]{\,\text{#1}}

\usepackage{accents} 
\newcommand\thickbar[1]{\accentset{\rule{.4em}{.8pt}}{#1}} 

\definecolor{redT}{RGB}{230,0,0}
\newcommand{\bSquare}{{\color{blue}$\square$}}
\newcommand{\bDiamond}{{\color{blue}$\Diamond$}}
\newcommand{\rTriangle}{{\color{redT}$\triangle$}}
\newcommand{\rTriangleDown}{{\color{redT}$\triangledown$}}

\hypersetup{
  colorlinks = true,
  urlcolor = blue,
  pdfauthor = {Kevin Slagle, Wonjune Choi, Li Ern Chern, Yong Baek Kim},
  pdftitle = {Slagle et al. - 2017 - Theory of a quantum spin liquid in hydrogen-intercalated honeycomb iridate, H3LiIr2O6}
}

\begin{document}

\title{Theory of a quantum spin liquid in hydrogen-intercalated honeycomb iridate, H${}_3$LiIr${}_2$O${}_6$}

\author{Kevin Slagle}
\affiliation{Department of Physics, University of Toronto, Toronto,
  Ontario M5S 1A7, Canada}
\author{Wonjune Choi}
\affiliation{Department of Physics, University of Toronto, Toronto,
  Ontario M5S 1A7, Canada}
  \author{Li Ern Chern}
\affiliation{Department of Physics, University of Toronto, Toronto,
  Ontario M5S 1A7, Canada}
\author{Yong Baek Kim}
\affiliation{Department of Physics, University of Toronto, Toronto,
  Ontario M5S 1A7, Canada}
\affiliation{Canadian Institute for Advanced Research, Toronto, Ontario, M5G 1M1, Canada}

\begin{abstract}
We propose a theoretical model for a gapless spin liquid phase that may have been
observed in a recent experiment on $\mathrm{H_3Li Ir_2 O_6}$ \cite{Takagi,TakagiTalk}.
Despite the insulating and non-magnetic nature of the material, the specific heat coefficient $C/T \sim 1/\sqrt{T}$ 
in zero magnetic field and $C/T \sim T/ B^{3/2}$ with finite magnetic field $B$ have been observed. 
In addition, the NMR relaxation rate shows $1/(T_1T) \sim (C/T)^2$.
Motivated by the fact that the interlayer/in-plane lattice parameters are
reduced/elongated by the hydrogen-intercalation of the parent compound $\mathrm{Li_2 Ir O_3}$,
we consider four layers of the Kitaev honeycomb lattice model with additional interlayer exchange 
interactions. It is shown that the resulting spin liquid excitations reside mostly in the top and bottom
layers of such a layered structure and possess a quartic dispersion. In an applied magnetic field, 
each quartic mode is split into four Majorana cones with the velocity $v \sim B^{3/4}$.
We suggest that the spin liquid phase in these ``defect" layers, 
placed between different stacking patterns of the honeycomb
layers, can explain the major phenomenology of the experiment, which can be taken as evidence that the Kitaev 
interaction plays the primary role in the formation of a quantum spin liquid in this material.
\end{abstract}

\maketitle

The honeycomb iridates $\mathrm{A_2 Ir O_3}$ (A=Na, Li) \cite{Trebst_2017,RauReview,YBKReview} have gained much attention \cite{Valenti2016,Valenti_2017,YJKim2013,YJKim2013b,Perkins2014,Perkins2015,You2011,Imada2014,Imada2016,Imada2017,Gegenwart2010,Cao2012,Coldea2012,Cao_Schlottmann_2017,BalentsSpinGap} as quantum spin liquid (QSL) candidate realizations of
  Kitaev's exactly solvable honeycomb lattice model \cite{KITAEV20062,CompassModels,Moessner2014}.
Due to crystal field splitting and spin-orbit coupling,
  the strongly correlated $5d$ electrons residing on the iridium ions can be described by an effective $j_{\rm eff} = \frac{1}{2}$ spin,
  and the bond-dependent Ising interactions of the Kitaev model can be realized due to a superexchange path through edge-shared oxygen octahedra \cite{JKmechanism}.
Although the Kitaev model has a spin liquid ground state,
  Kitaev materials such as $\mathrm{Na_2IrO_3}$ or $\alpha$-$\mathrm{Li_2IrO_3}$ are magnetically ordered at low temperatures \cite{Cao2012,PhysRevB.93.195158,ZigzagOrder}.
This occurs because an additional exchange path from a direct overlap of iridium orbitals introduces additional
  Heisenberg \cite{PhysRevLett.105.027204} and anisotropic off-diagonal exchange ($\Gamma$) \cite{PhysRevLett.112.077204} interactions,
  which favor a magnetically ordered ground state. 

In a recent experiment, Takagi and his colleagues have synthesized a new spin liquid candidate material, $\mathrm{H_3Li Ir_2 O_6}$,
  by substituting the interlayer lithium ions of $\alpha$-$\mathrm{Li_2 Ir O_3}$ by hydrogen.
This insulating material shows no sign of magnetic order down to low temperatures in the magnetic susceptibility, specific heat, and NMR measurements \cite{Takagi,TakagiTalk},
raising the hope for discovery of a quantum spin liquid.
The X-ray powder diffraction pattern suggests a heavily stacking-faulted crystal structure with an enlarged in-plane bond length and 
reduced interlayer distance.
The longer in-plane bond length can be expected to suppress the Heisenberg and anisotropic off-diagonal exchange ($\Gamma$) interactions since
  the contribution from direct exchange is greatly reduced,
  which can allow the Kitaev interaction to dominate the physics.

The experiment is especially significant since the material may be the first material that is a Kitaev-like spin liquid;
  the first to be \emph{engineered} to be a spin liquid;
  and the first where strong inter-layer coupling stabilizes a spin liquid.
Furthermore, given the close connection to Kitaev's exact solution,
  the candidate spin liquid has a strong theoretical foundation.
This also suggests that when a magnetic field is applied,
  the material could be in an Ising topologically ordered phase with nonabelian anyons \cite{KITAEV20062}
  relevant to fault-tolerant quantum computation \cite{KITAEV20032}.

However, the NMR spin relaxation rate $1/T_1$ and the specific heat $C$ disagree with thermodynamic properties of a pure Kitaev spin liquid with Majorana cones
  (for which $C/T \sim T$).
Instead, it is found that
\begin{equation}
  (T_1T)^{-1/2} \sim C/T \sim T^{-1/2} \label{eq:C}
\end{equation}
at low temperatures ($0.06\unit{K} < T < 2\unit{K}$), which implies an abundant density of states at low energies.
But in the presence of an external magnetic field $B$ (with $1\unit{Tesla} \leq B \leq 8\unit{Tesla}$ and temperature $0.1\unit{K} \lesssim T \lesssim 1\unit{K}$),
\begin{equation}
  (T_1T)^{-1/2} \sim C/T \sim B^{-3/2} T \label{eq:C B}
\end{equation}
In the experiment, the magnetic entropy obtained by integrating the specific heat data suggests 
that only a few percent of the local moments contribute to the singular specific heat.
This suggest that the specific heat may be dominated by 
unusual ``defects'' in the material. \cite{TakagiTalk}

In this paper, we propose a theoretical model for a gapless spin liquid that may explain
these experiments.
Because the interlayer distance is shortened,
  we expect interlayer interactions to play important roles.
Thus, in addition to the Kitaev in-plane interaction,
  we introduce interlayer exchange interactions to couple the Kitaev honeycomb layers.
We assume that there is a small fraction of ABCA-type stacked layers in the crystal (to be generalized later);
  e.g., the complete sequence could contain ($\dots$B[ABCA]C$\dots$).
Since the lattice and the stacking patterns are very likely distorted 
  from the ideal structure, we consider the distortion effect via further neighbor 
  exchange interactions instead of taking into account the distortion of the lattice itself.

Similar to ABC-stacked multilayered graphene \cite{PhysRevB.77.155416,PhysRevB.81.125304},
  we show that a coupled ABCA-stack of Kitaev spin liquids has Majorana excitations with a quartic dispersion \cite{footK4}.
As shown later, these soft modes are mostly localized in the top and bottom layers and hence represent
  two-dimensional states.
The density of states due to this four-layer ``defect'' stacking pattern is given by $D(E) \sim E^{-1/2}$,
  which explains the spin relaxation rate and specific heat (before magnetic fields are applied) in \eqnref{eq:C}.
With only a small number of ABCA-type stacking between
different stacking patterns,
the magnetic entropy due to these ``defect" layers contributes
only a small fraction of the total entropy, as seen in
the experiment.
In the presence of a magnetic field, each quartic mode is split into four Majorana cones in our model.
The momentum shift (from the quartic touching point) $k_0$ of the Majorana cones scales as $k_0 \sim B^{1/4}$
  since the energy shift is $\Delta E \sim k^4_0 \sim B$ with the Zeeman coupling.
Therefore the velocity of the Majorana fermions is $v \sim k_0^3 \sim B^{3/4}$,
  and the Majorana cones (in two spatial dimensions) have a density of states \cite{Foot:DoS}
\begin{equation}
  D(E) \sim E/v^2 \sim B^{-3/2} E
  \label{eq:DoS}
\end{equation}
  which produces the scaling in \eqnref{eq:C B}.
Similar to Kitaev's exactly solvable model \cite{KITAEV20062},
  a small gap $E_0 \sim 10^{-3}\unit{meV}$ can be expected \cite{Foot:coneGap},
  which may only be observable at significantly lower temperatures ($\sim 0.01\unit{K}$).

\section{Model}
\begin{figure}
\begin{minipage}{.49\columnwidth}
\includegraphics[width=\textwidth]{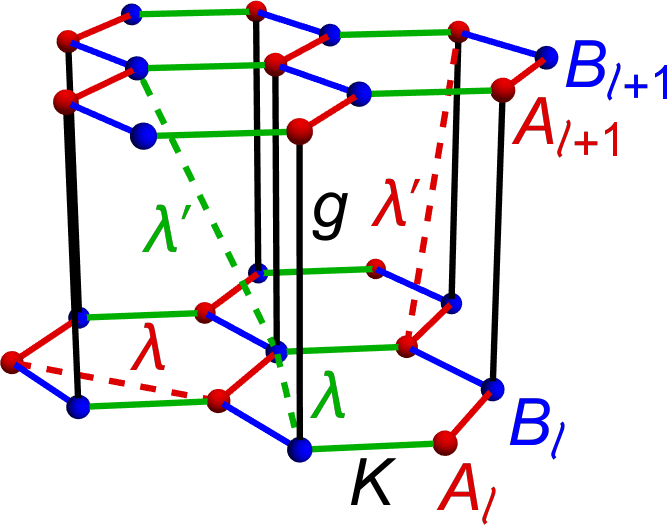}
\textbf{(a)}
\end{minipage}
\begin{minipage}{.49\columnwidth}
\includegraphics[width=\textwidth]{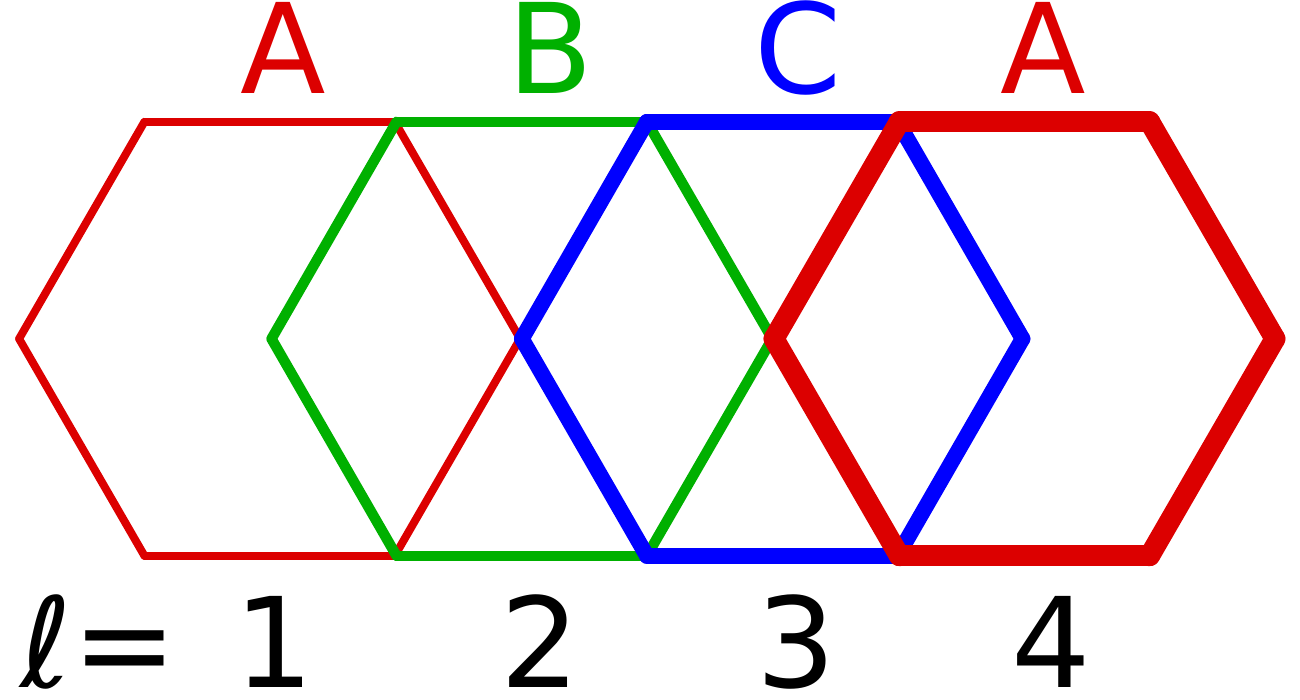}
\textbf{(b)}
\end{minipage}
\caption{
\textbf{(a)}
Two of the four layers in our model (\eqnref{eq:H}).
Red and blue vertices denote the A and B sublattices, respectively.
The red, green, and blue links correspond to $\sigma^x\sigma^x$, $\sigma^y\sigma^y$, and $\sigma^z\sigma^z$ couplings, respectively.
The solid colored links denote Kitaev couplings in $H_K$ and are summed over by $\sum_{\langle i,j \rangle \in \mu}$.
The black links denote interlayer Heisenberg couplings in $H_g$ and are summed by $\sum_{\langle i,j \rangle}^{\text{A}\bar{\text{B}}}$.
The dotted red and green links denote the $\sigma^x\sigma^x$ and $\sigma^y\sigma^y$ couplings, respectively, that appear in in $H_\lambda$ and $H_{\lambda'}$.
Note that the dotted couplings are highly anisotropic;
  all of the dotted couplings for one unit cell have been drawn.
(A unit cell has two sites per layer. For drawing clarity, some of the $\lambda$ and $\lambda'$ couplings have been translated into neighboring unit cells.)
\textbf{(b)}
A hexagon from each of the four layers ($\ell=1,2,3,4$) when viewed directly from above,
  which demonstrates what is meant by ABCA stacking.
}\label{fig:lattice}
\end{figure}

\begin{figure}
\includegraphics[width=.95\columnwidth]{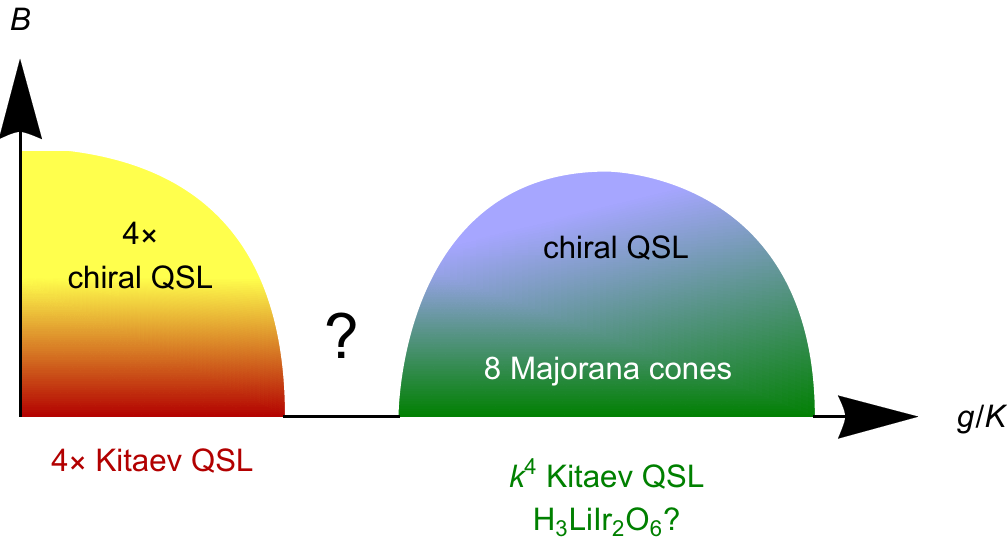}
\caption{
Phase diagram of our model (\eqnref{eq:H}).
\textbf{(red)} When $g/K$ is small and $B=0$,
  our model is in the same phase as four decoupled layers of Kitaev's QSL honeycomb model \cite{KITAEV20062},
  where each layer can be described by two gapless Majorana cones coupled to a $Z_2$ gauge field \cite{Foot:LayerStability}.
\textbf{(yellow)} However, a magnetic field ($B$) opens up a small gap and the resulting phase is four copies of a chiral QSL \cite{Foot:coneGap}. 
\textbf{(green)} According to mean-field theory, for intermediate $g/K$ and $B=0$,
  our model is described by two Majorana modes with quartic dispersion \cite{footK4} coupled to a $Z_2$ gauge field.
\textbf{(green$\rightarrow$blue)}
When a small magnetic field ($B$) is applied,
  each of the two Majorana modes with quartic dispersion split into four Majorana cones (eight in total) with linear dispersion (\sfigref{fig:dispersion}{a}).
However, our model actually predicts a very small gap (see \sfigref{fig:dispersion}{b}) for these Majorana cones \cite{footQSL}.
\textbf{(white)}
Contents of the white region are unknown.
}\label{fig:phases}
\end{figure}

The Hamiltonian that we consider consists of an ABCA-type stacking of $N=4$ honeycomb lattices.
Each honeycomb layer hosts a Kitaev honeycomb model \cite{KITAEV20062} described by $H_K$,
  and the layers are coupled together by a Heisenberg interaction ($H_g$) (\figref{fig:lattice}).
We will also consider additional in-plane ($H_\lambda$) and interlayer ($H_{\lambda'}$) interactions.
\begin{align}
  H   &= H_K + H_g + H_\lambda + H_{\lambda'} \label{eq:H}\\
  H_K &= K \sum_{\ell=1}^N \sum_{\mu=x,y,z} \sum_{\langle i,j \rangle \in \mu}
    \sigma_{\ell,i}^\mu \sigma_{\ell,j}^\mu \quad\quad,\; N=4 \nonumber\\
  H_g &= g \sum_{\ell=1}^{N-1} \sum_{\langle i,j \rangle}^{\text{A}\thickbar{\text{B}}}
    \vec\sigma_{\ell+1,i} \cdot \vec\sigma_{\ell,j} \nonumber\\
  H_\lambda    &= \lambda \sum_{\ell=1,N} \left(
    \sum_{\langle\langle i,j \rangle\rangle}^{\text{AA}_x} \sigma_{\ell,i}^x \sigma_{\ell,j}^x +
    \sum_{\langle\langle i,j \rangle\rangle}^{\text{BB}_y} \sigma_{\ell,i}^y \sigma_{\ell,j}^y \right) \label{eq:H lambda}\\
  H_{\lambda'} &= \lambda' \sum_{\ell=1,N-1} \left(
    \sum_{\langle\langle i,j \rangle\rangle}^{\text{A}\thickbar{\text{A}}_x} \sigma_{\ell+1,i}^x \sigma_{\ell,j}^x +
    \sum_{\langle\langle i,j \rangle\rangle}^{\text{B}\thickbar{\text{B}}_y} \sigma_{\ell+1,i}^y \sigma_{\ell,j}^y \right) \nonumber
\end{align}
The summations $\sum_{\langle i,j \rangle \in \mu}$, $\sum_{\langle i,j \rangle}^{\text{A}\thickbar{\text{B}}}$, and $\sum_{\langle\langle i,j \rangle\rangle}^{\cdots}$ sum over the pairs of lattice sites indicated in \sfigref{fig:lattice}{a}.
The magnitude of $K$ and $g$ are not known,
  but $K$ is likely to be similar to the value for its parent material $\alpha$-Li$_2$IrO$_3$ \cite{Valenti2016}, and $g$ could be similar:
  $-K \sim g \sim 10\unit{meV}$.
We will also couple the model to a magnetic field $B_\mu$
\begin{equation}
  H_B = - \sum_{\ell,i,\mu} B_\mu \sigma_{\ell,i}^\mu
\end{equation}
See \figref{fig:phases} for a mean-field phase diagram for this model.

Notice that $H_\lambda$ is a next-nearest neighbor, bond and sublattice dependent, intralayer, Ising coupling.
For our purposes, it will be sufficient to consider this interaction on the boundary layers, but it could also 
be present in every layer. $H_{\lambda'}$ is similar, except it is an interlayer coupling.
Later, we show that without $H_\lambda$ or $H_{\lambda'}$, our mean field model would
  result in a non-generic magnetic field dependence of the density of states (\eqnref{eq:DoS}).
$H_\lambda$ or $H_{\lambda'}$ are just two possible examples of how to obtain the observed generic magnetic field dependence (in \appref{app:general lambda} we consider more general possibilities);
  either alone is sufficient.
  The underlying lattice distortion may render the magnitude of their coupling 
  constants ($\lambda$, $\lambda'$) as large as the nearest neighbor coupling $g$.
  On the other hand, \eqnref{eq:DoS} will only hold for sufficiently small magnetic fields:
  $B \lesssim B_\text{max} \lesssim \max(\lambda,\lambda')$.

\begin{figure}
\includegraphics[width=.7\columnwidth]{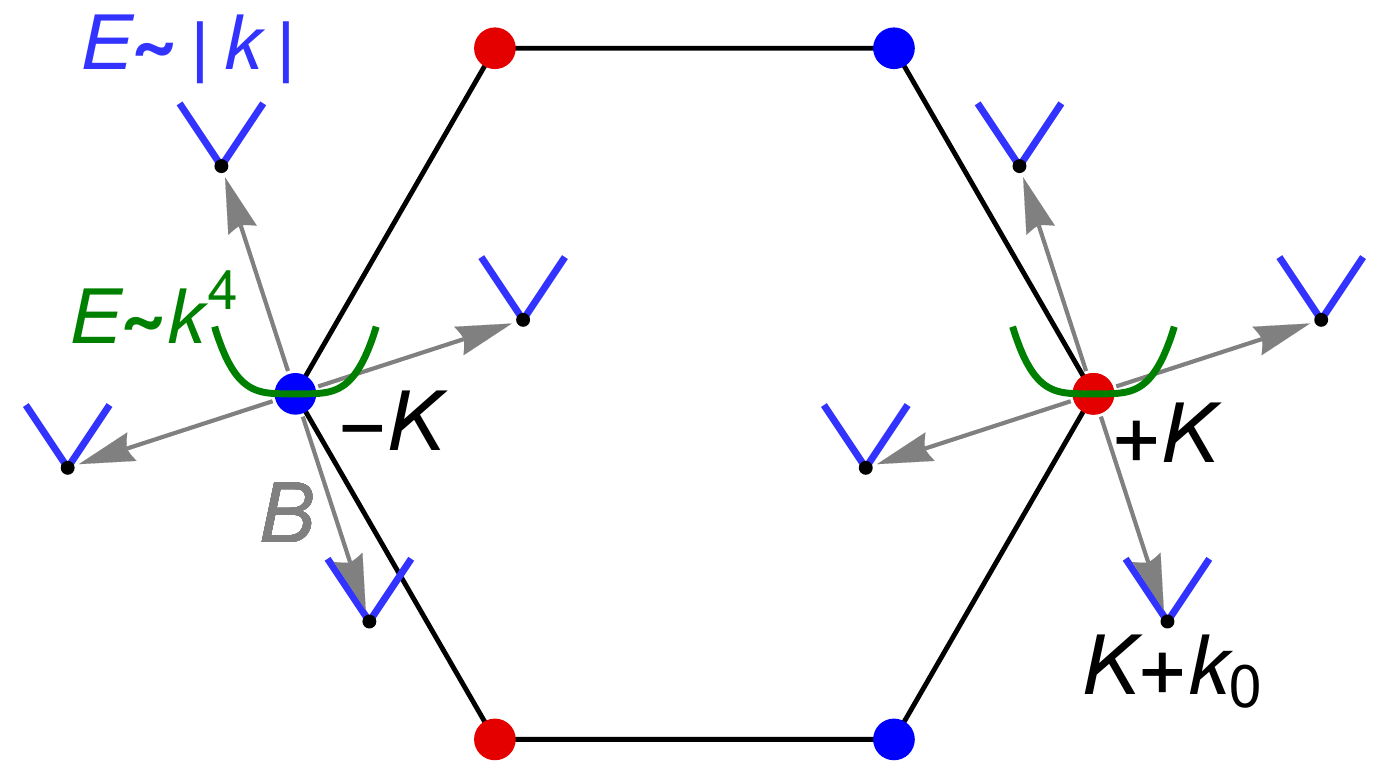} \\
\textbf{(a)} Brillouin zone \\
\includegraphics[width=.7\columnwidth]{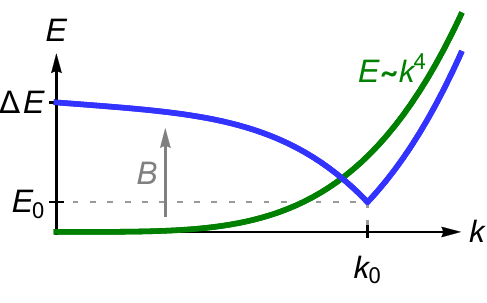} \\
\textbf{(b)} dispersion
\caption{
\textbf{(a)}
Before a magnetic field ($B$) is applied, our model has two gapless Majorana modes with quartic dispersion (green curve) at the $\pm K$ points (red and blue dots).
(Note that in the Brillouin zone, the three red dots are equivalent points.)
After a $B$ field is applied, the quartic mode splits into four Majorana cones (blue cones) which are displaced by momentum $|k_0|$.
\textbf{(b)}
The dispersion of the Majorana fermions along one of the gray arrows in (a).
\textbf{(green)} Quartic dispersion before a magnetic field ($B$) is applied.
\textbf{(blue)} Majorana cone after a $B$ field is applied.
$\Delta E \sim B$, $|k_0| \sim B^{1/4}$, and $E_0 \sim B^3$ \cite{Foot:coneGap}.
See \figref{fig:detailedDispersion} in the appendix for more detailed plots.
}\label{fig:dispersion}
\end{figure}

\section{Mean-Field Theory}

We now study our model using mean-field theory.
We follow Kitaev and decompose the spins into four Majorana fermions \cite{KITAEV20062}:
  $\sigma_{\ell i}^\mu = \ii\, b_{\ell i}^\mu c_{\ell i}$. 
The physical states ($\ket{\psi}$) must obey the following Hilbert space constraint:
  $b_{\ell i}^x b_{\ell i}^y b_{\ell i}^z c_{\ell i} \ket{\psi} = \ket{\psi}$. 

After decomposing the spins,
  all of the terms in our Hamiltonian become products of four Majorana fermions.
We will apply mean-field theory in order to obtain a solvable quadratic Hamiltonian.
For $H_K$ and $H_g$,
  we will use the mean-field decomposition:
\begin{align}
\begin{split}
  \sigma_{\ell i}^\mu \sigma_{\ell' j}^\mu &\stackrel{\text{MF}}{\approx}
    - \langle \ii\, b_{\ell i}^\mu b_{\ell' j}^\mu \rangle         \ii\, c_{\ell  i}     c_{\ell' j}
    - \langle \ii\, c_{\ell i}     c_{\ell' j}     \rangle         \ii\, b_{\ell  i}^\mu b_{\ell' j}^\mu \label{eq:MF Kg}\\
    &\quad\;\; + \langle \ii\, b_{\ell i}^\mu b_{\ell' j}^\mu \rangle \langle \ii\, c_{\ell  i}     c_{\ell' j} \rangle
\end{split}
\end{align}
If we only consider the Kitaev's honeycomb model $H_K$,
  then this approximation is exact since it reproduces Kitaev's exact solution \cite{KITAEV20062}.
The approximation is also exact if we consider only the Heisenberg Hamiltonian $H_g$
  in the sense that it results in the expected dimerized ground state (of spin singlet pairs across the Heisenberg bonds)
  after projecting into the physical Hilbert space. 
Thus, we expect this decomposition to be accurate in the colored regions of our phase diagram (\figref{fig:phases}).


After inserting the mean-field decompositions, we
  Fourier transform the Majorana fermions:
\begin{equation}
  \begin{pmatrix}
    c_{k\ell\alpha} \\ b_{k\ell\alpha}^\mu
  \end{pmatrix}
  = \sum_{i \in \alpha} e^{-\ii\, (K+k) \cdot i}
  \begin{pmatrix}
    c_{\ell i} \\ b_{\ell i}^\mu
  \end{pmatrix} \label{eq:Fourier}
\end{equation}
where $\alpha$ ($=A,B$) is the sublattice of site $i$. \cite{Foot:fermions}
$\pm K$ are the locations of the gapless points (\sfigref{fig:dispersion}{a}) so that $k$ is the momentum displacement from these points.
Since we are only interested in the low energy physics,
  we will expand about small $k$.
Finally, we
  rotate the phase of the $c$ and $b$ fermions on the $B$ and $A$ sublattices in order to cancel out factors of $\ii$ in $H^\text{MF}$;
  i.e. $c_{k\ell B} \rightarrow \ii c_{k\ell B}$ and $b_{k\ell A}^\mu \rightarrow -\ii b_{k\ell A}^\mu$.

\begin{figure}
\includegraphics[width=\columnwidth]{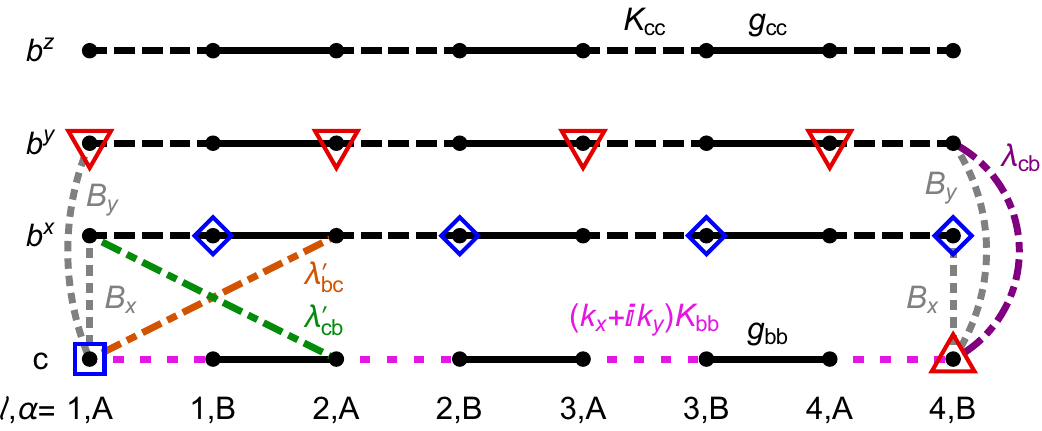}
\caption{
Picture of our mean-field Hamiltonian (\eqnref{eq:H MF}).
The single-particle Hamiltonian of the $c$-fermions at a given momentum $k$
  resembles a fermion SPT chain \cite{KitaevChain} with low-energy modes at the ends of the chain
  (corresponding to the top and bottom layers).
When the momentum $k$ is shifted away from the gapless points (\sfigref{fig:dispersion}{a}),
  the correlation length of the SPT chain increases and the energy of the edge modes is $k^N$,
  where $N$ ($=4$ above) is the length of the chain.
Please see paragraphs below \eqnref{eq:H MF} for further explanation.
}\label{fig:kspace}
\end{figure}

The mean-field Hamiltonian (which is depicted in \figref{fig:kspace}) then takes the form of $H^\text{MF} = \int_k H^\text{MF}_k$ where:
\begin{align}
  H^\text{MF}_k &=
      K_\text{bb} \sum_{\ell=1}^N (k_x + \ii k_y) c_{k\ell A}^\dag c_{k\ell B}
    + g_\text{bb} \sum_{\ell=1}^{N-1} c_{k,\ell+1,A}^\dag c_{k\ell B} \nonumber\\
   &- K_\text{cc} \sum_{\ell,\mu} e^{\ii \phi_\mu} b_{k\ell A}^{\mu\dag} b_{k\ell B}^\mu
    + g_\text{cc} \sum_{\ell,\mu} b_{k,\ell+1,A}^{\mu\dag} b_{k\ell B}^\mu \label{eq:H MF}\\
   &- \sum_{\ell,i,\mu} B_\mu \ii\, b_{k\ell i}^{\mu\dag} c_{k\ell i}
    + h.c. + H^\text{MF}_{\lambda;k} \nonumber\\
   & (\phi_x,\phi_y,\phi_z) = (-1,1,3) \; \pi \ii/3 \label{eq:phi}
\end{align}
For simplicity, we absorbed the mean-field amplitudes into the coupling constants
(e.g. $K_\text{bb}        \equiv   K        \langle \ii\, b_{\ell i}^\mu    b_{\ell j}^\mu \rangle$). 
We will ignore $H^\text{MF}_{\lambda;k}$ until later.

Since $H^\text{MF}$ is quadratic and translation invariant,
  each momentum component decouples.
$H^\text{MF}_k$ is composed of $4 \times (N=4) \times 2 = 32$ (complex) fermion operators,
  each denoted by a black dot in \figref{fig:kspace}:
  the $4$ flavors ($c,b^x,b^y,b^z$) are positioned along the rows
  while the $N=4$ layers ($\ell$) and two ($\alpha=A,B$) sublattices form the columns.
Thus, for a given momentum $k$,
  we can picture $H^\text{MF}_k$ as four chains of complex fermions.
The $g_\text{cc}$ and $g_\text{bb}$ terms in $H^\text{MF}_k$ couple the fermions connected by the solid black lines,
  and $K_\text{cc}$ couples $b$ fermions connected by the dashed black lines.
We will consider $g_\text{cc} \sim g_\text{bb} \sim K_\text{cc} \sim K_\text{bb} \sim 1$.
With only these terms (i.e. $k=B=0$),
  the $c$ fermions at the ends of the chain (\bSquare\ and \rTriangle) are decoupled and form zero energy eigenstates.
When a small $k$ is introduced,
  a small $(k_x + \ii k_y) K_\text{bb}$ term couples the $c$ fermions across the dotted pink lines.
The $c$ fermion chain then resembles a fermion chain symmetry protected topological (SPT) model \cite{KitaevChain},
  where the edge modes have a gap that is exponentially small in the length ($2N$) of the chain: $E \sim k^N$.
Since $N=4$, we see that $H^\text{MF}_k$ has a quartic dispersion,
  which leads to the specific heat in \eqnref{eq:C}.

A magnetic field $B$ couples the $c$ and $b$ fermions:
  i.e. $B_\mu$ couples each $c$ fermion to the $b^\mu$ above it in \figref{fig:kspace}.
Four examples of $B_\mu$ are shown in \figref{fig:kspace} as dotted gray lines.
Although the $c$ fermion chain is an SPT with a very short correlation length (when $k$ is small),
  the $b$ fermion chain is gapped with a correlation length comparable to the length of the chain (when $g_\text{cc} \sim K_\text{cc}$).
Thus, a small magnetic field perturbation will couple the $c$ fermion edge modes (\bSquare\ and \rTriangle) at second order in perturbation theory since
  a fermion at \bSquare\ will have to hop across two magnetic field perturbations (and accross the four \bDiamond\ or \rTriangleDown\ in \figref{fig:kspace})
  in order to get to \rTriangle.
An effective Hamiltonian describing the low energy $c$ fermion edge modes will thus include a term with energy coefficient $\sim B^2$.
When we back out of the spin chain picture and think about what happens to the quartic dispersion,
  we find that it actually spits into $N=4$ Dirac cones, shifted by momenta $|k_0| \sim B^{1/2}$ with 
  the velocity $v \sim |k_0|^3 \sim B^{3/2}$ and density of states $D(E) \sim E/v^2 \sim B^{-3} E$.
However, this scaling is not generic;
  it occurs because the magnetic field only contributed at second order in perturbation theory,
  which resulted because our mean-field model was fine-tuned such that the $c$ and $b$ fermions do not mix.

In order to mix the $c$ and $b$ fermions,
  we need to introduce an additional term in our Hamiltonian.
As an example of how this mixing could occur, we consider
  $H_\lambda$ and $H_{\lambda'}$ (\eqnref{eq:H lambda}) with the following mean-field decomposition:
\begin{align}
\begin{split}
  \sigma_{\ell i}^\mu \sigma_{\ell' j}^\mu &\stackrel{\text{MF}}{\approx}
    + \langle \ii\, b_{\ell i}^\mu c_{\ell' j}     \rangle         \ii\, c_{\ell  i}     b_{\ell' j}^\mu
    + \langle \ii\, c_{\ell i}     b_{\ell' j}^\mu \rangle         \ii\, b_{\ell  i}^\mu c_{\ell' j}     \label{eq:MF lambda}\\
   &\quad\;\; - \langle \ii\, b_{\ell i}^\mu c_{\ell' j}     \rangle \langle \ii\, c_{\ell  i}     b_{\ell' j}^\mu \rangle
\end{split}
\end{align}
This results in the following additional terms to the mean-field Hamiltonian (\eqnref{eq:H MF}):
\begin{align}
   H^\text{MF}_{\lambda;k}
   &= \lambda_\text{cb} \sum_{\ell=1,N} \left(
          b_{k\ell A}^{x\dag} c_{k\ell A}
        - b_{k\ell B}^{y\dag} c_{k\ell B} \right) \label{eq:H MF lambda}\\
   &+\lambda'_\text{cb} \sum_{L=1,N-1} \Big[ e^{\ii \phi_x} \left(
                        b_{k,\ell+1,A}^{x\dag} c_{k\ell A}
                      + c_{k,\ell+1,A}^  \dag  b_{k\ell A}^x \right) \nonumber\\
   &\hspace{2cm} + e^{\ii \phi_y} \left(
                        b_{k,\ell+1,B}^{y\dag} c_{k\ell B}
                      + c_{k,\ell+1,B}^  \dag  b_{k\ell B}^y \right) \Big] \nonumber
\end{align}
These terms couple the $c$ fermions on the $A$ sublattice to the $b^x$ fermions,
  and the $c$ fermions on the $B$ sublattice to the $b^y$ fermions.
A few examples of these couplings are drawn in \figref{fig:kspace}.
If $\lambda_\text{cb} \neq 0$,
  then the $b^y$ fermions and the $c$ fermion at \rTriangle\ form a chain of length $9$,
  and the eigenvector with \rTriangle\ now also includes contributions from \rTriangleDown\ with amplitude
  $\psi_0 \sim \max(\lambda_\text{cb}, \lambda'_\text{cb})$
  (when $\max(\lambda_\text{cb}, \lambda'_\text{cb}) \lesssim g_\text{cc} \sim K_\text{cc} \sim 1$).
This eigenstate is still a zero mode since the length of the chain is odd.
The physics is the same if we consider $\lambda'$ terms instead.
Similarly, the \bSquare\ eigenstate includes contributions from \bDiamond with the same amplitude $\psi_0$.
This is important since now the two zero modes (with support over \bSquare\bDiamond\ or \rTriangle\rTriangleDown)
  are directly coupled by the magnetic field $B$ (via the dotted gray lines shown in \figref{fig:kspace}).
Thus, following the logic of the previous paragraph,
  the $B$ field now enters at first order in perturbation theory and introduces a term with energy coefficient $\Delta E \sim B$
  to the effective Hamiltonian describing the low energy modes.
The $B$ field now splits the quartic mode into $N=4$ Dirac cones shifted by momenta $k_0 \sim B^{1/4}$, with 
velocity $v \sim B^{3/4}$ and
density of states $D(E) \sim B^{-3/2} E$.
This is precisely the scaling seen in the experiment \cite{Foot:DoS}.

\section{Discussion}

Motivated by a recent experiment on $\mathrm{H_3Li Ir_2 O_6}$ \cite{Takagi,TakagiTalk}, 
we have proposed a model for a quantum spin liquid in coupled-layers of Kitaev spin liquids.
We use an example of ABCA-type stacked-layers of the Kitaev spin liquid
with the nearest and next-nearest interlayer interactions, which were 
used to mimic the effect of lattice distortion in real material.
In the mean-field theory, 
we show that the scaling of the specific heat and NMR relaxation rate seen 
in the experiment can be explained by the 
underlying gapless Majorana fermions, which are localized near the top and bottom layers of the coupled-layer
system.

On phenomenological ground, we are assuming that the ABCA-type stacking
pattern makes up a small fraction of the possibe stacking patterns that 
may exist in $\mathrm{H_3Li Ir_2 O_6}$. The singular specific heat contribution 
from such ``defect'' patterns will be a small portion of the total magnetic entropy,
which is consistent with the specific heat data.
While the spin susceptibility in the presence of strong spin-orbit coupling
does not simply reflect the density of states of spinful excitations,
the bulk susceptibility, which includes the contributions from
the ``defect" layers, is related to the specific heat via a thermodynamic
relation. This is clearly demonstrated in the experiment. \cite{Takagi,TakagiTalk}
In contrast, the Knight shift shows very little temperature 
dependence at low temperatures, which may be consistent with the 
expectation that the Knight shift is relatively insensitive to those ``defects".
Going beyond mean-field theory, a small magnetic field opens a small mass gap for the Majorana cones \cite{Foot:coneGap}.
However, it may be difficult to see such a small gap in the
  experimental regime of $T \sim 0.1-1$K and $B \sim$ 1-8 T, where the
  characteristic scalings of $1/T_1$ and $C/T$ were observed;
  smaller temperatures and larger magnetic fields may be needed.
As shown in the case of stacked graphene layers \cite{PhysRevB.77.155416,PhysRevB.81.125304},
  there exist other multi-layer stacking patterns
  where soft modes with quartic dispersion exists (e.g. ABCAC or ABCACB) (along with other less-soft modes),
  or cubic $k^3$ (ABC) or quintic $k^5$ (ABCAB).
As such, other $\mathrm{H_3Li Ir_2 O_6}$ samples could also exhibit different dispersions which are dominated by various kinds of stacking sequences.
Further experiments on the distribution of the stacking patterns could be helpful.


\begin{acknowledgments}
We thank H. Takagi and K. Kitagawa for sharing their experimental data and for helpful discussions.
This work was supported by the NSERC of Canada 
and the Center for Quantum Materials at the University of Toronto.
We acknowledge the hospitality at the Kavli Institute for Theoretical Physics, supported in part by 
the NSF Grant No. PHY-1125915 and
the Aspen Center for Physics, supported in part by NSF Grant No. PHY-1607611,
where some parts of this work were done.
\end{acknowledgments}

\bibliography{k4qsl16}

\begin{thebibliography}{38}
\bibitem [{foo()}]{footK4}%
  \BibitemOpen
  \href@noop {} {}\bibinfo {note} {We expect that additional generic (but weak)
  interactions can split each quartic mode into four Majorana cones. However,
  very low temperatures may be required to observe this effect.}\BibitemShut
  {Stop}%
\bibitem [{Foo({\natexlab{a}})}]{Foot:DoS}%
  \BibitemOpen
  \href@noop {} {} ({\natexlab{a}}),\ \bibinfo {note} {{I}n general, gapless
  fermions (or bosons) with a $E \sim v k^\alpha$ dispersion in $d$ spatial
  dimensions has a density of states $D(E) \sim (E/v)^{d/\alpha} E^{-1}$, which
  can be roughly observed in experiments as the ratio of specific heat over
  temperature: $C/T \sim D(T)$.}\BibitemShut {Stop}%
\bibitem [{Foo({\natexlab{b}})}]{Foot:coneGap}%
  \BibitemOpen
  \href@noop {} {} ({\natexlab{b}}),\ \bibinfo {note} {{K}itaev showed that
  $E_0 \sim B^3/K^2$ (see \sfigref{fig:dispersion}{b}) using perturbation
  thoery \cite{KITAEV20062}. In the material \cite{Takagi,TakagiTalk}, one
  could estimate $E_0 \sim (8\unit{$\mu_e$T})^3 / (10\unit{meV})^2 \sim
  10^{-3}\unit{meV} \sim 0.01\unit{k$_\text{B}$K}$, which is well-below the
  lowest temperature observed in the experiment ($0.1\unit{K}$). Additional
  terms, such as an intralayer anisotropic off-diagonal exchange ($\Gamma$),
  could increase the gap to $E_0 \sim \Gamma B/K$. But $\Gamma$ and thus $E_0$
  can still be expected to be small due to the enlarged in-plane bond
  length.}\BibitemShut {Stop}%
\bibitem [{Foo({\natexlab{c}})}]{Foot:LayerStability}%
  \BibitemOpen
  \href@noop {} {} ({\natexlab{c}}),\ \bibinfo {note} {{A} stack of Kitaev
  honeycomb models should be stable to a weak Heisenberg coupling because this
  coupling preserves time reversal. This can be understood by following the
  arguments in \refcite{KITAEV20062}. A perturbation expansion relative to the
  gauge sector of the ground state results in an effective Hamiltonian with
  interaction terms coupling the layers of the form $c_{\ell i} c_{\ell j}
  c_{\ell'i'} c_{\ell'j'}$ (neglecting $\ii\, b^\mu_{\ell i} b^\mu_{\ell j}$
  factors). However, quartic interactions are an irrelevant (in the RG sense)
  perturbation to gapless Majorana cones in two spatial dimensions, and thus
  the phase is stable.}\BibitemShut {Stop}%
\bibitem [{foo()}]{footQSL}%
  \BibitemOpen
  \href@noop {} {}\bibinfo {note} {{D}ue to the small gap, the blue phase in
  \figref{fig:phases} is topologically ordered with Chern number $\pm 4$, which
  we calculated using Eq.\,54 in Ref.\,\cite{KITAEV20062}.}\BibitemShut {Stop}%
\bibitem [{Foo()}]{Foot:fermions}%
  \BibitemOpen
  \href@noop {} {}\bibinfo {note} {Note that although $c_{\ell i}$ and $b_{\ell
  i}^\mu$ in \eqnref{eq:Fourier} are Majorana fermions, $c_{k\ell\alpha}$ and
  $b_{k\ell\alpha}^\mu$ are complex fermions. As a result, the gapless point of
  the complex fermion $c_{k\ell\alpha}$ at $k=0$ describes the two gapless
  points $\pm K$ of the Majorana fermions. $c_{k\ell\alpha}$ and
  $b_{k\ell\alpha}^\mu$ are complex fermions because we applied a momentum
  displacement $K$ to their definition. As a result, $c_{k\ell\alpha} \neq
  c_{k\ell\alpha}^\dag \neq c_{-k,\ell,\alpha}$. (If instead $K=0$, then
  $c_{k\ell\alpha}$ would be Majorana since then $c_{k\ell\alpha}^\dag =
  c_{-k,\ell,\alpha}$.) In this context, the difference between Majorana and
  complex fermions is purely mathematical: two Majorana modes (zero modes,
  Dirac cones, or quartic dispersing modes) is equivalent to a single complex
  fermion mode.}\BibitemShut {Stop}%
\end{thebibliography}

\appendix

\begin{figure*}
\centering
\begin{minipage}{.95\columnwidth}
  $B=0$
  \includegraphics[width=\textwidth]{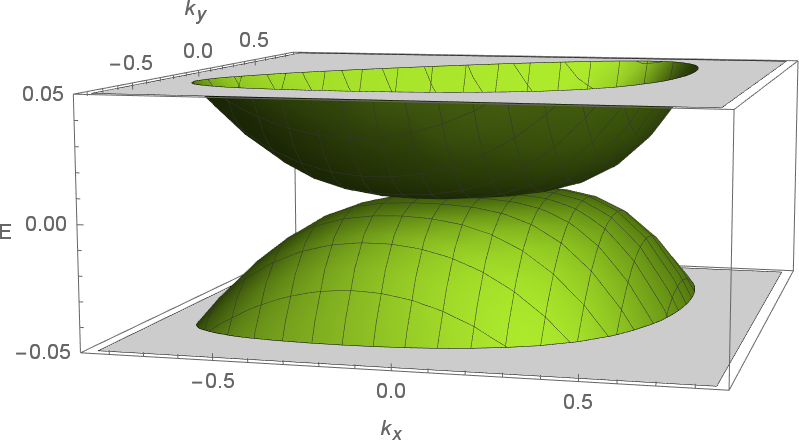}
\end{minipage}
\hspace{.08\columnwidth}
\begin{minipage}{.95\columnwidth}
  $B_\mu = 1/16$
  \includegraphics[width=\textwidth]{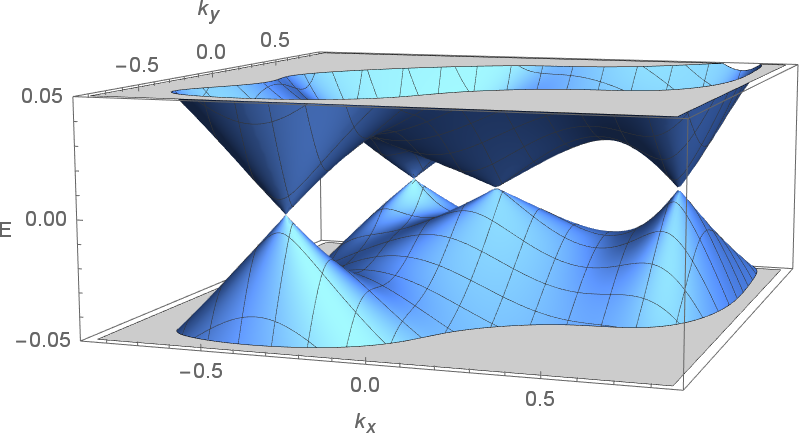}
\end{minipage} \\
\begin{minipage}{.95\columnwidth}
  \includegraphics[width=\textwidth]{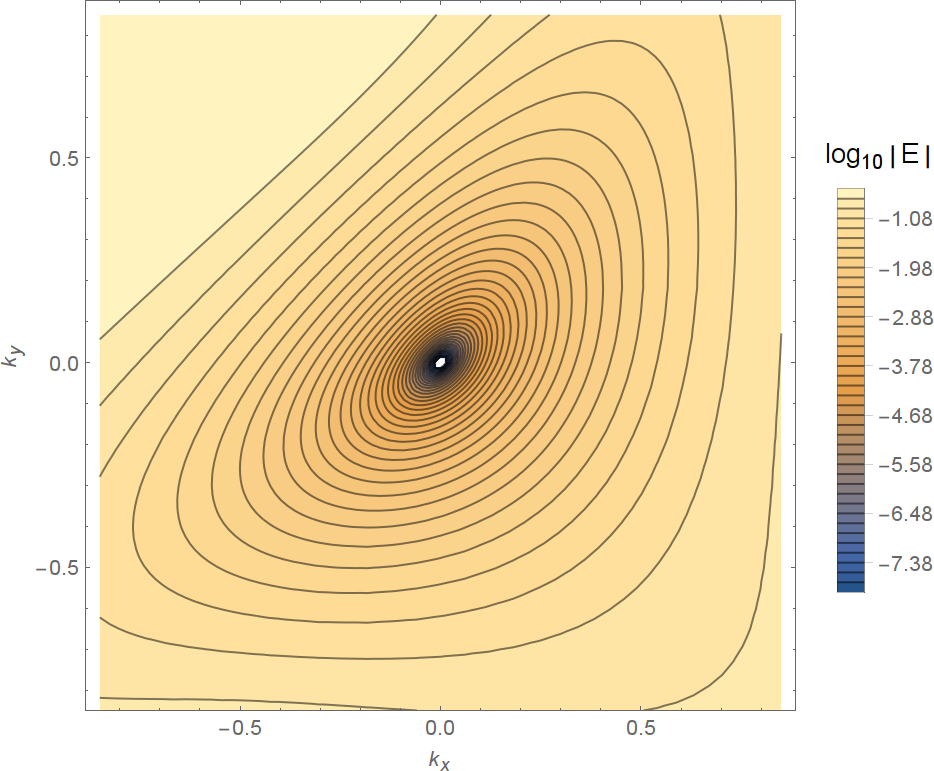}
\end{minipage}
\hspace{.08\columnwidth}
\begin{minipage}{.95\columnwidth}
  \includegraphics[width=\textwidth]{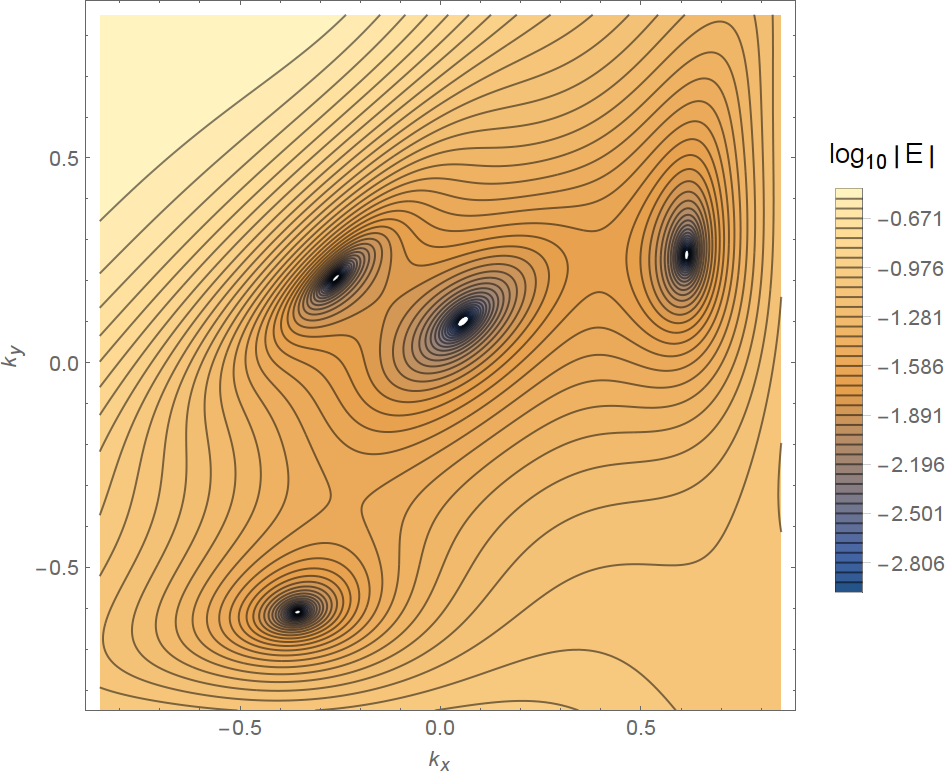}
\end{minipage}
\caption{
Example low energy band structure in units where $K=g=1$ and with $\lambda=\lambda'=1/4$ near a $K$ point in the Brillouin zone.
The left column is for no magnetic field, while the right column includes a small magnetic field $B_\mu = 1/16$ for $\mu=x,y,z$.
The bottom row is a contour plot of $\log_{10} E$, where $E$ is the band energy.
In the above plots, we are not including the extra terms that Kitaev generated via perturbation theory in the presence of a magnetic field \cite{Foot:coneGap},
  which is why no gap is present in the above plots.
}\label{fig:detailedDispersion}
\end{figure*}

\section{Other \texorpdfstring{$\lambda$}{lambda} Terms}
\label{app:general lambda}

In \eqnref{eq:H lambda} we considered a simple possible example for $H_\lambda$ and $H_{\lambda'}$.
Here, we will explain a more general example:
\begin{align}
  H_\lambda    &= \sum_{\ell\alpha\mu} \sum_{\langle\langle i \in (\ell,\alpha), j \in (\ell  ,\alpha) \rangle\rangle \perp \mu}
                    \lambda_{\ell\alpha\mu\nu}  \sigma_{\ell    i}^\nu \sigma_{\ell j}^\nu \\
  H_{\lambda'} &= \sum_{\ell\alpha\mu} \sum_{\langle\langle i \in (\ell,\alpha), j \in (\ell+1,\alpha) \rangle\rangle \in   \mu}
                    \lambda'_{\ell\alpha\mu\nu} \sigma_{\ell+1, i}^\nu \sigma_{\ell j}^\nu \nonumber
\end{align}
$\sum_{\langle\langle i \in (\ell,\alpha), j \in (\ell,\alpha) \rangle\rangle \perp \mu}$
  sums over all pairs of sites $(i,j)$ where $i$ is on layer $\ell$ and sublattice $\alpha$,
  and similarly $j \in (\ell,\alpha)$,
  and where $\mu$ specifies the direction of the $(i,j)$ bond.


This choice contributes to $H^\text{MF}_k$ (\eqnref{eq:H MF}) as follows:
\begin{align}
  H^\text{MF}_k &= \sum_{\ell\alpha\nu} \left(B_\nu + \sum_\mu \lambda_{\ell\alpha\mu\nu} \right) b_{k\ell\alpha}^{\nu\dag} c_{k\ell\alpha} \\
    &\quad - \left(\sum_\mu \lambda'_{\ell\alpha\mu\nu} \, e^{\ii \phi_\mu}\right) \left(
      b_{k,\ell+1,\alpha}^{\nu\dag} c_{k\ell\alpha} - c_{k,\ell+1,\alpha}^\dag b_{k\ell\alpha}^\nu \right) \nonumber\\
    &\quad + h.c. + \cdots \nonumber
\end{align}
where $\phi_\mu$ was defined in \eqnref{eq:phi}.
We see that $\lambda'$ must depend on the bond direction $\mu$,
  or else it cancels out above.

However, there are other constraints that must be imposed on $\lambda$ and $\lambda'$,
  which can be understood from \figref{fig:kspace}.
In particular, if the $A$ sublattice has a $\sigma^\nu \sigma^\nu$ coupling,
  then the $B$ sublattice must not also have this coupling.
That is,
\begin{align}
  \text{if } \lambda_{\ell A\mu\nu} \neq 0 \text{ or } \lambda'_{\ell A\mu\nu} \neq 0, \nonumber\\
  \text{then } \lambda_{\ell B\mu\nu} \approx \lambda'_{\ell B\mu\nu} \approx 0 \label{eq:lambda constraint}
\end{align}
and similar for $A \leftrightarrow B$.
If the above is not true, e.g. if $\lambda_{\ell\alpha\mu\nu} = \lambda'_{\ell\alpha\mu\nu} = 1$,
  then even before a magnetic field is applied,
  the zero modes (\bSquare\ and \rTriangle) would be coupled to each other,
  which would split the quartic dispersion into Majorana cones.
However, in a material, all of these $\lambda$ terms can be expected to be nonzero.
But most of them will probably be very small;
  and as long as \eqnref{eq:lambda constraint} is at least approximately obeyed,
  a quartic dispersion will be observed in the specific heat until a very low temperature,
  which has not been observed yet.

\end{document}